 \documentclass[manuscript]{emulateapj}

 \usepackage{apjfonts}

\slugcomment{to appear in the Astrophysical Journal}

\shorttitle{Light Curve Model of M31N 2008-12a}
\shortauthors{Kato et al.}

\begin{document}


% should be written in CAPITAL

\title{MULTI-WAVELENGTH LIGHT CURVE MODEL OF THE ONE-YEAR RECURRENCE
PERIOD NOVA M31N 2008-12A}


\author{Mariko Kato} 
\affil{Department of Astronomy, Keio University, Hiyoshi, Yokohama
  223-8521, Japan;}
\email{mariko.kato@hc.st.keio.ac.jp}

\author{Hideyuki Saio}
\affil{Astronomical Institute, Graduate School of Science,
    Tohoku University, Sendai 980-8578, Japan}
% \email{saio@astr.tohoku.ac.jp}

\and
\author{Izumi Hachisu}
\affil{Department of Earth Science and Astronomy, College of Arts and
Sciences, The University of Tokyo, 3-8-1 Komaba, Meguro-ku,
Tokyo 153-8902, Japan}
%\email{hachisu@ea.c.u-tokyo.ac.jp}



\begin{abstract} 
We present a theoretical light curve model of the recurrent nova 
M31N 2008-12a, the current record holder for the shortest recurrence 
period (1 yr).
We combined interior structures calculated using a Henyey-type
evolution code with optically thick wind solutions of hydrogen-rich
envelopes, which give the proper mass-loss rates,
photospheric temperatures, and luminosities. 
The light curve model is calculated for a 1.38 $M_\sun$ white dwarf (WD) 
with an accretion rate of $1.6 \times 10^{-7}~M_\sun$~yr$^{-1}$. 
This model shows a very high effective temperature 
($\log T_{\rm ph}$ (K) $\geq 4.97$) and a very small wind mass-loss rate
($\dot M_{\rm wind} \leq 9.3 \times 10^{-6}~M_\sun$~yr$^{-1}$)
even at the maximum expansion of the photosphere. 
These properties are consistent with the faint optical peak 
of M31N 2008-12a because the brightness of the free--free emission is 
proportional to the square of the mass-loss rate.
The model well reproduces the short supersoft X-ray turn-on time of 6 days and 
turnoff time of 18 days after the outburst. 
The ejecta mass of our model is calculated to be $6.3 \times 10^{-8}~M_\sun$, 
corresponding to 37\% of the accreted mass.   
The growth rate of the WD is 0.63 times the mass accretion rate, 
making it a progenitor for a Type Ia supernova. 
Our light curve model predicts a bright supersoft X-ray phase one or two days 
before the optical peak. 
We encourage detection of this X-ray flash in future outbursts. 
\end{abstract}

% up to 6 keywords

\section{INTRODUCTION} \label{sec_introduction}

A nova is a thermonuclear runaway event on a mass-accreting white dwarf (WD) 
in a binary. When hydrogen burning sets in,  
the envelope of the WD greatly expands, and the 
WD becomes bright in the optical band. 
Wind mass-loss begins, carrying away part of the envelope mass. 
After the optical maximum, the photosphere moves inward, and  
the main emitting wavelength region of photons shifts to
a shorter wavelength region.   
Subsequently, the WD becomes a supersoft X-ray source (SSS), 
and the SSS phase continues until the end of hydrogen burning. 

The recent discovery of the $\sim 1$ yr recurrence period nova M31N 2008-12a
has drawn attention to novae of short recurrence periods 
\citep{dar14,dar15, hen14,hen15, tan14}. 
Recurrent novae are binaries harboring a massive WD. 
One-year recurrence periods occur for very massive WDs of
$M_{\rm WD}\gtrsim 1.3~ M_\sun$ and very high mass accretion rates of
$\dot M_{\rm acc}\gtrsim1.5 \times 10^{-7}M_\sun$~yr$^{-1}$ 
\citep{pri95,wol13,tan14,kat14shn}. 
Such massive WDs are considered to be among the candidates for 
Type Ia supernova (SN~Ia) progenitors
\citep{hku99, hkn99, hac01kb, hkn10, han04, li97, kat12review}. 
SNe Ia play an  important role in astrophysics as a standard candle  
for measuring cosmological distances, and as the main producers of iron
group elements in the chemical evolution of galaxies.  
However, their immediate progenitors remain elusive. The current debate centers on the single degenerate (SD) versus double degenerate (DD) scenarios
\citep[e.g., ][]{mao14,pag14}. 
In SD scenarios \citep[e.g., ][]{hku99,hkn99,kat12review},  
very short recurrence periods in novae indicate the final stages 
before an SN Ia explosion.
The recurrent nova M31N 2008-12a has a recurrence period of one year, the 
shortest on record, and is thus considered to be an immediate progenitor.     
The detailed properties of an immediate progenitor could shed new light on the debate between DD and SD scenarios. 
Thus, theoretical and observational studies of M31N 2008-12a 
are very important.

In the last outburst of M31N 2008-12a in October 2014, unprecedented observational data were
obtained, including the optical rising phase, 
X-ray turn-on time ($t_{\rm on}$), and X-ray turnoff time ($t_{\rm off}$) 
\citep{dar15,hen15}.
The nova exhibits a faint optical peak, which is 1--2 mag fainter 
than the galactic recurrent novae
U Sco, whose recurrence time is $t_{\rm rec}=$8--20 yr and RS Oph 
whose recurrence time is $t_{\rm rec}= $ 10--20 yr, and has 
a very short X-ray turn-on time 
($t_{\rm on}=5.9 \pm 0.5$ days after the outburst),
followed by a supersoft X-ray phase of 12 days with a high effective
temperature of up to $120$ eV  (a blackbody fit of the {\it Swift} XRT 
spectrum by \citet{hen15}).
The X-ray turnoff time is 
$t_{\rm off}=18.4 \pm 0.5$ days.  
All of these characteristics indicate a very massive WD.

In Section \ref{section_SSSduration},  
we estimate the WD mass from the duration of an SSS phase.  
Our light curve model is presented in Section \ref{section_results}. 
Discussion and conclusions follow in Section \ref{sec_discussion}.

\section{DURATION OF SUPERSOFT X-RAY PHASE}\label{section_SSSduration}

\begin{figure}
\epsscale{0.9}
%% \plotone{Xrayduration.eps}
\plotone{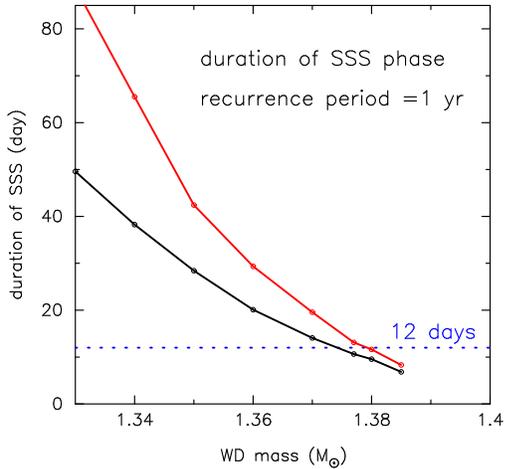}
\caption{
Duration of supersoft X-ray source (SSS) phase of one-year-recurrence
period novae. The black line represents no accretion in the SSS phase, 
i.e., accretion resumed after $t_{\rm off}$ time.
The red line shows the case in which the accretion resumed
at $t_{\rm on}$ time. 
The accretion rate is taken from Kato et al.'s (2014) calculation 
for one-year recurrence period novae, 
e.g., $1.6 \times 10^{-7}~M_\sun$~yr$^{-1}$ for 1.38~$M_\sun$, 
$2.5 \times 10^{-7}~M_\sun$~yr$^{-1}$ for 1.35~$M_\sun$, 
and $3.3 \times 10^{-7}~M_\sun$~yr$^{-1}$ for 1.33~$M_\sun$.
The blue horizontal dotted line corresponds to 
the SSS duration (12 days) of the recurrent nova M31N 2008-12a. 
\label{SSS.duration}}
% source: m31n2008-12a/Xrayduration.wip
\end{figure}

In an SSS phase, the envelope on the WD is geometrically thin,
and in  hydrostatic balance.  This phase is well represented
by a sequence of hydrostatic solutions \citep[e.g.,][]{kat94h,sal05}.
We calculated the duration of the SSS phase for the WD masses,
1.33, 1.34, 1.35, 1.36, 1.37, 1.377, 1.38, and 1.385 $M_\sun$. 
The chemical composition of the envelope is assumed to be uniform in space 
and constant in time, but to have a linear dependence on the WD mass, 
i.e., $X = 0.6$ for $1.33~M_\sun$, and $X=0.55$ for $1.385~M_\sun$
with $Z=0.02$, the value of which are taken from \citet{kat14shn}. 
We also include the so-called gravitational energy release,
which amounts to 10\% of the photospheric luminosity. 
Figure \ref{SSS.duration} shows the duration of the SSS phase, 
with the time between the X-ray turn-on and turnoff times  
defined as the epoch when the optically thick winds stop, 
and the epoch when hydrogen burning is extinguished, respectively \citep{kat94h,hac10}.

If the accretion disk is not destroyed because of very weak winds, 
the accretion may resume soon after the winds stop.   
In this case, we have a longer duration of the SSS phase 
because additional nuclear fuel (hydrogen) is being supplied. 
The red line in Figure \ref{SSS.duration} represents 
the SSS duration in the presence of accretion, the rates of which 
are taken from those of one-year recurrence period novae \citep{kat14shn},  
 e.g., $1.6 \times 10^{-7}~M_\sun$~yr$^{-1}$ for 1.38~$M_\sun$,
$2.5 \times 10^{-7}~M_\sun$~yr$^{-1}$ for 1.35~$M_\sun$,
and $3.3 \times 10^{-7}~M_\sun$~yr$^{-1}$ for 1.33~$M_\sun$.
  
This accretion effect is more significant in the 1.33 $M_\sun$ WD
than in more massive WDs because the accretion rate is close to
the steady hydrogen burning rate in the 1.33 $M_\sun$ WD, 
but much smaller in the 1.385 $M_\sun$ WD. 
From this figure, we conclude that the nova M31N 2008-12a
($t_{\rm SSS} \sim 12$ days) harbors a very massive WD,
(as massive as $1.38~M_\sun$) in  cases both 
with and without accretion in the SSS phase.

% Fig.1
%\placefigure{SSS.duration}

\section{LIGHT CURVE MODEL}\label{section_results}

We calculated nova outbursts on the $1.38~M_\sun$ WD  
accreting hydrogen-rich matter ($X=0.7,~Y=0.28$, and $Z=0.02$
for hydrogen, helium, and heavy elements, respectively)
at a rate of $1.6 \times 10^{-7}~M_\sun$ yr$^{-1}$. 
We calculated eight cycles until the flash reaches the limit cycle  
 using the same Henyey-type code as used by \citet{kat14shn}. 
The recurrence period of flashes is 1.07 yr. 
During the shell flash, the photospheric radius (as well as the luminosity) 
increases, and the temperature decreases beyond the peak of OPAL opacities 
\citep[$\log T$ (K) $\sim 5.2$,][]{igl96}. This triggers wind mass loss. 
In such a stage, we adopt the optically thick wind solutions \citep{kat94h} 
as the surface boundary condition.  The wide band 
spectrum--energy distribution of the 2014 outburst of M31N 2008-12a 
showed flat spectra \citep[see Figure 11 of][]{dar15}.
This means that the spectrum is of free--free emission. 
The optical and ultraviolet (UV) light curves are calculated using 
the wind mass-loss rate, photospheric radius, and velocity, assuming 
free--free emission \citep[Equation (2) in][]{hac08kc,hac10}.
The absolute flux of the free--free emission is determined 
from the distance and extinction to M31. 
After the optically thick winds stop, the envelope structure approaches 
that of hydrostatic balance. 
The X-ray light curve is calculated assuming a blackbody emission for 
the photospheric temperature and luminosity of the envelope.  
We stopped the accretion during the wind phase and restarted it
at the beginning of the SSS phase.

%Fig.2
%\placefigure{light}

\begin{figure}
%\epsscale{0.9}
\epsscale{1.1}
%% \plotone{light.opal10.f.eps}
\plotone{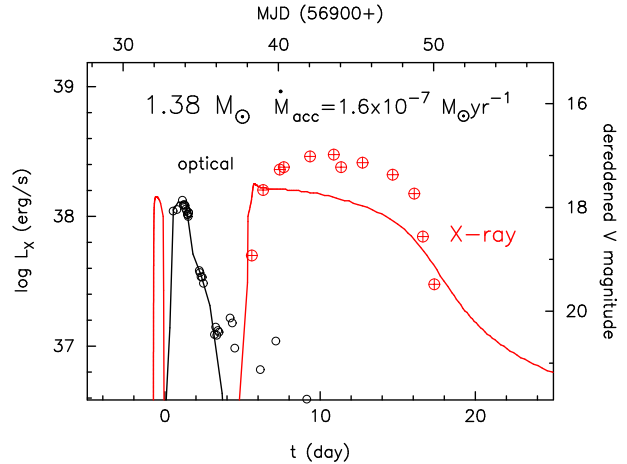}
\caption{
Multi-wavelength light curve model of M31N 2008-12a. 
The optical (black circles) and 
X-ray data (red encircled pluses) are taken from \citet{dar15} and 
\citet{hen15}, respectively. 
The optical data are dereddened with $A_{\rm V}=0.64$.  
The solid red lines represent our theoretical supersoft
X-ray (0.2--1 keV) light curve.  
The black solid line denotes our theoretical 
free--free emission light curve for the V band. 
See text for more detail. 
\label{light}}
% source: m31n2008-12a/light.opal10.wip 
\end{figure}

Figure \ref{light} shows our multi-wavelength light curves.  
The solid red lines represent the theoretical supersoft X-ray flux (0.2--1 keV). 
In the very early phase of the outburst, the photospheric temperature
increases, keeping the radius almost constant up to 
$\log T_{\rm ph}$ (K) = 6.1,  then decreases with expansion.
Thus, we have a short X-ray bright phase (X-ray flash) one or two days 
before the optical peak.  

After the envelope expands, the optically thick wind begins to blow, 
and lasts for 5.4 days.  The black line in Figure \ref{light}
is the optical brightness calculated on the basis of free--free emission. 
The brightness reaches its maximum at the maximum expansion of the 
photosphere, when the photospheric temperature reaches its minimum, 
although as high as $\log T_{\rm ph}$ (K) = 4.97, 
 the wind mass-loss rate attains
its maximum at $\dot M_{\rm wind}=9.3 \times 10^{-6}~M_\sun$ yr$^{-1}$. 
This mass-loss rate is much smaller than that of typical classical novae 
\citep{kat94h}, which explains the faint optical peak of M31N 2008-12a 
\citep[See Figure 36 in ][]{hac15k}
because the brightness of the free--free emission is
proportional to the square of the mass-loss rate.  
The internal structure of the winds is essentially the same as in 
\citet{kat94h}. 
After the optical maximum, the flux of free--free emission 
decays quickly as the wind-mass loss rate decreases.  

Our optical light curve well reproduces
the observed $V$ magnitude light curve in the first three days, as 
shown in Figure \ref{light}, where the data are taken
from Table 6 in \citet{dar15}.  
We assumed an absorption of $A_{\rm V}= R_{\rm V} \times E(B-V) =  
3.1 \times 0.21 = 0.64$, where we adopt the hydrogen column density 
$N_{\rm H}=1.4 \times 10^{21}$ cm$^{-2}$ \citep{hen14,hen15} and 
the relationship $E(B-V)=N_{\rm H}/6.8 \times 10^{21}$ cm$^{-2}$ \citep{gue09}. 

After day 3, the observed optical magnitude decays much more slowly than 
the theoretical light curve. 
These excess fluxes could not come from either the free--free emission, because 
the wind has already stopped, or from the photospheric blackbody emission,
because its temperature is too high to emit at the optical bands. 
They could be an indication of the beginning of a nebular phase,
that is, the contribution of strong emission lines \citep[see, e.g., ][]{hac10} 
or of a plateau phase, as observed in other recurrent novae, 
i.e., emission from an irradiated disk,
such as in U~Sco \citep{hkkm00} and RS Oph \citep{hac06a}.

After day 5, the photospheric temperature 
increases enough to emit X-rays, and the nova enters an SSS phase. 
Our theoretical light curve of the supersoft X-ray band well reproduces 
the observed X-ray turn-on and turnoff times.  
The theoretical temperature rises to 
$\log T_{\rm ph}$ (K) = 6.16 (i.e., 125 eV), 
 consistent with $kT \sim120$ eV obtained from a blackbody fit
of the {\it Swift} XRT spectrum \citep{hen15}.

\section{DISCUSSION AND CONCLUSIONS}\label{sec_discussion}

Our light curve model predicts a bright X-ray phase before the optical maximum. 
In our $1.38~M_\sun$ model, the X-ray flash begins 
at $t=-0.73$ day (i.e., 1.85 days before the optical peak at $t=1.12$ days) 
and lasts 0.64 days, as shown in Figure \ref{light}. Here, we regard the X-ray phase as the time the X-ray luminosity spends above $3.0 \times 10^{37}$ erg~s$^{-1}$ 
(based on the blackbody luminosities in \citet{hen15}). 
The evolution time is longer in less massive WDs, and so is the duration 
of the X-ray flash. 
For a $1.35~M_\sun$ WD with a mass accretion rate of $2.5 \times 10^{-7}~M_\sun$~yr$^{-1}$, 
the X-ray flash starts 7.0 days before the optical maximum, and lasts 1.4 days. 
The timescale may also depend on other quantities, such as 
the mass accretion rate and chemical composition of the accreted matter. 
However, our estimate of an X-ray flash of $\sim 1$ day duration starting $\sim 2$ days 
before the optical maximum is robust unless the duration of the SSS phase after 
the optical peak is changing.

The mass lost in the wind phase is calculated to be 
$6.3 \times 10^{-8}~M_\sun$. 
This value is very small compared with those of typical classical novae 
($\sim 10^{-5}$--$10^{-4}~M_\sun$). 
The accreted matter is $ 1.7 \times 10^{-7}~M_\sun$. 
Thus, the lost mass amounts to 37\% of the accreted matter. 
The mass accumulation efficiency is $\eta =0.63$ in our 1.38~$M_\sun$ WD. 

In our calculation, the ejecta is hydrogen-rich, with $X \sim 0.6$. 
This small decrease in $X$ from the initial value of $X=0.7$ to $X=0.6$  
is caused by convective mixing in the very early phase of the shell flash. 
\citet{hen15} estimated the ejected hydrogen mass to be 
$M_{\rm ej,H}=(2.6 \pm 0.4)\times 10^{-8}~M_\sun$. 
Considering the hydrogen content of $X \sim 0.6$, our value of 
($ 6.3 \times 10^{-8}~M_\sun \times 0.6 = 3.8 \times 10^{-8}~M_\sun$)
is roughly consistent with the observational estimate. 
 \citet{dar15} derived a total ejected mass of $\gtrsim 3 \times 10^{-8} M_\sun$ 
based on the observed deceleration of the ejecta and a comparison to RS Oph. 
Our value is also consistent with this estimate.

%Fig.3
%\placefigure{light.uv}

\begin{figure}
%\epsscale{0.9}
\epsscale{1.1}
%% \plotone{light.opal10.uv.eps}
\plotone{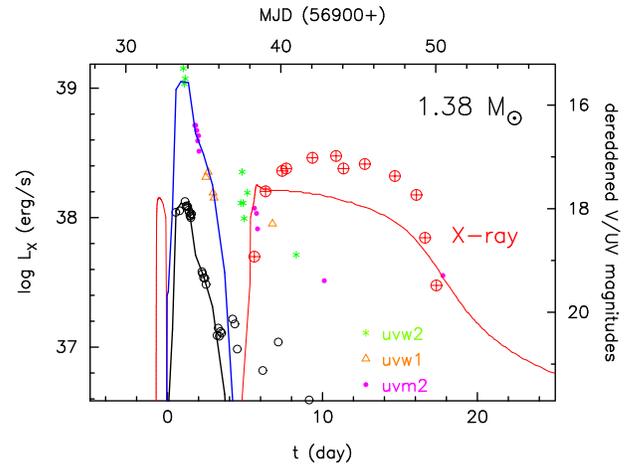}
\caption{
Same as Figure \ref{light}, but with added UV light curves. 
The UV data of {\it Swift} UVOT bands, $uvw2$ (green asterisks),
$uvw1$ (orange open triangles), and $uvm2$ (magenta filled circles)
are taken from \citet{dar15}. They  
are all shifted upward by the same amount of $A_{\rm UV}=1.71$, 
instead of being corrected by dereddening. 
The blue line is our theoretical free--free emission light curve
for the UV wavelength bands. 
See text for more detail. 
\label{light.uv}}
% source: m31n2008-12a/light.opal10.wip 
\end{figure}

M31N 2008-12a was very bright in the UV bands in the optically bright phase,  
as shown in Figure \ref{light.uv}. 
We suggest that these UV fluxes are free--free emission from the ejecta 
just outside the photosphere, based on the following.  
The first reason is the wide-band spectrum--energy--distribution 
\citep{dar15}, which is 
consistent with free--free emission, not only in the V band,
but also in these UV bands.   The second reason is the light curve shape.  
The blue line in Figure \ref{light.uv} has the same shape 
as the black line, but is shifted up by 2.35 mag. This blue line shows 
good agreement with the UV data, although the three UV bands, $uvm2$,
$uvw1$, and $uvw2$, have different band-passes.
If free--free emission dominates the spectrum, these light curves should
have the same shape \citep[e.g.,][]{hac06kb,hac15k}. 
This agreement strongly indicates that the UV emission is free--free emission.
The third reason is that these strong UV fluxes cannot be explained by 
 photospheric blackbody emission, which is as small as
$L_{\rm max} = 1.4 \times 10^{37}$ erg s$^{-1}$
for the $uvw2$ band at $\log T_{\rm ph}$ (K) $=  4.97$ (maximum expansion
of the photosphere). 
In future outbursts, we encourage high cadence UV observation 
to confirm the wavelength-independent shape
of the light curve.

Our main results are summarized as follows.

\begin{enumerate}
\item
We calculated the duration of the supersoft X-ray phase of novae of 
one-year recurrence periods for various WD masses. 
The duration of 12 days suggests that the  
recurrent nova M31N 2008-12a harbors a WD as massive as $ \sim 1.38~M_\sun$. 
This value is close to the 
upper limit of the mass-accreting WD because the mass-accreting WDs have a hot core,  
and the upper limit is less than the Chandrasekhar mass limit \citep[e.g., ][]{nom82}. 
\item
We modeled the M31N 2008-12a outbursts as a nova outburst 
on a 1.38 $M_\sun$ WD with an accretion rate of 
$1.6 \times 10^{-7}~M_\sun$ yr$^{-1}$.  
This model explains the optical/UV light curves on the basis of free--free emission
originating from winds, as well as the 
supersoft X-ray light curve on the basis of blackbody emission from the WD photosphere. 
The smaller wind mass-loss rates and higher photospheric temperatures 
are consistent with the faint optical peak of M31N 2008-12a. 
\item
The ejected mass is calculated to be $\sim 6 \times 10^{-8}~M_\sun$.
This corresponds to the mass accumulation efficiency of $\eta =0.63$. 
Thus, the WD is increasing in mass. 
This makes M31N 2008-12a a strong candidate for an SN Ia progenitor. 
\item
The ejected mass is much smaller than typical classical novae 
($\sim 10^{-5}$--$10^{-4}~M_\sun$),  but is roughly consistent with
observational estimates of ejected total mass of 
 $\gtrsim 3 \times 10^{-8} M_\sun$ \citep{dar15} and ejected 
hydrogen mass of $2.6 \times 10^{-8}~M_\sun$ \citep{hen15}.  
\item
The next outburst of the recurrent nova M31N 2008-12a is expected 
in autumn 2015.  
We encourage detection of this X-ray flash  
in order to obtain a complete description of this valuable outburst.
Also, detection of a UV flux during the X-ray flash would give us unprecedented
information on the irradiated accretion disk or circumbinary matter before 
it is disturbed by the high velocity ejecta. 
High cadence observation of  UV light curves, as well as spectra, 
would be very useful for studying the origin of these emissions.

\end{enumerate}

\acknowledgments
We are grateful to Martin Henze for fruitful discussions. We also thank the 
anonymous referee for useful comments that improved the manuscript. 
This research was supported in part by the
Grants-in-Aid for Scientific Research (15K05026 and 24540227)
from the Japan Society for the Promotion of Science.

%todayreference
%  for more than 5 authors: only first three authors, + et al.
%

%todayfigure

%Fig.1 

%Fig.2  

%Fig.3  

\end{document}